\documentclass[prl,preprint,showpacs,superscriptaddress]{revtex4}
\usepackage{graphicx}
\usepackage{dcolumn}

\def\gdot{\dot{\gamma}}

\def\mob{\mbox{\sffamily\bfseries M}}
\def\be{\begin{equation}}
\def\beq{\begin{equation}}
\def\ee{\end{equation}}
\def\eeq{\end{equation}}
\def\bea{\begin{eqnarray}}

\def\eea{\end{eqnarray}}
\newcommand{\pder}[2]{\frac{\displaystyle\partial#1}{\displaystyle\partial#2}}
\begin{document}


 \title{THEORY OF SUSPENSION SEGREGATION IN PARTIALLY FILLED
HORIZONTAL ROTATING CYLINDERS}
\author{Rama Govindarajan}
\email{rama@jncasr.ac.in}
\affiliation{$^1$Fluid Dynamics Unit,
Jawaharlal Nehru Centre for Advanced Scientific
Research, Jakkur, Bangalore 560 064, India} 
\author{Prabhu R. Nott}
\email{prnott@chemeng.iisc.ernet.in}
\affiliation{Department of Chemical Engineering, 
Indian Institute of Science, Bangalore 560 012, India}  
\author{Sriram Ramaswamy\footnote{Also with JNCASR, Bangalore 560 064 India}}
\email{sriram@physics.iisc.ernet.in}
\affiliation{Centre for Condensed-Matter Theory, Department of Physics,
Indian Institute of Science, Bangalore 560 012, India}  

\begin{abstract}
\baselineskip 12pt
It is shown that a suspension of particles in a partially-filled,
horizontal, rotating cylinder is linearly unstable towards axial
segregation and an undulation of the free surface at large enough
particle concentrations. Relying on the shear-induced diffusion of
particles, concentration-dependent viscosity, and the existence of
a free surface, our theory provides an explanation of the
experiments of Tirumkudulu {\it et al.}, Phys. Fluids {\bf 11},
507-509 (1999); {\it ibid.} {\bf 12}, 1615 (2000).
\end{abstract}

\maketitle

\begin{figure}
\begin{center}
\includegraphics{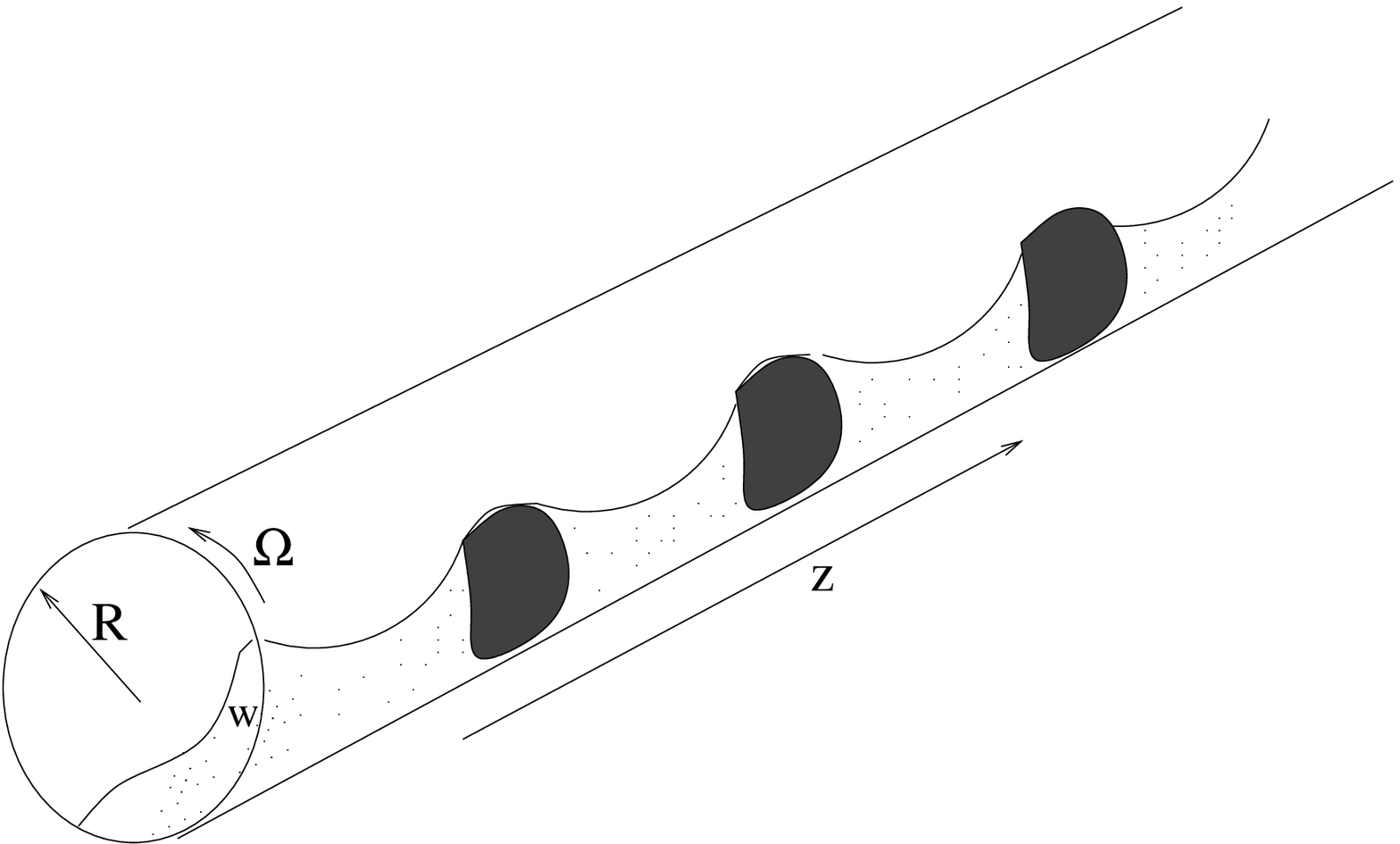}
\end{center}
\caption{\label{fig:cyl} 
}
\end{figure}

\baselineskip 16pt
The primary motivation for this work is to understand the
experiments of \cite{acr1,tiru}, showing spontaneous segregation
in sheared suspensions. In these experiments, monodisperse polymer
spheres several hundred microns in diameter were suspended
 uniformly in fluids about 1000 times as viscous as water,
with the same mass density as the particles, so that there was no
sedimentation. The initial experiments \cite{acr1} were carried
out with a suspension in a horizontal Couette cell, i.e., in the
gap between two concentric cylinders, but in subsequent studies
\cite{tiru}, a single horizontal cylinder of radius $R$ was used. In both sets
of experiments the container was filled only partially, i.e.,
there was a free surface. Let us restrict our description to the
experiments in \cite{tiru} for simplicity. When the cylinder was
rotated at a tangential speed $v_0 = R \Omega$ about its symmetry
axis the initially uniform suspension was found to undergo a
dramatic instability towards segregation into bands of high and
low concentration, with wavevector along the cylinder axis. The
surface profile was modulated as well, i.e., the thickness of the
fluid layer varied along the axis (see Fig. \ref{fig:cyl}),
with thicker regions corresponding to higher concentration. No
instability was seen if there was no free surface, i.e.~when the
cylinder was completely filled with suspension.

The phenomena reported in the experiments \cite{acr1,tiru} are
among the many intriguing effects known to arise in suspensions of
non-Brownian particles in highly viscous fluids, driven by shear
flow or sedimentation. The feature of these suspensions that is of
relevance in this paper is that the particles diffuse \cite{leigh}
even though their thermal Brownian motion is negligible. The
microscopic explanation \cite{leigh,bossis,janosi} for this
diffusion is that the hydrodynamic interaction between the
particles renders their motion chaotic, even in the Stokesian
limit where inertia of fluid and particles is ignored. The
 diffusive flux of particles has two parts, one driven by a
 gradient in the particle volume fraction $\phi$, the other by a
 gradient in the deformation-rate $\gdot$.  For situations in which the
 predominant variation is with respect to a single coordinate $z$
(e.g., the axial coordinate  of the cylinder in \cite{acr1,tiru})
and time $t$ the local volume fraction $\phi(z, t)$ (integrated
over the remaining directions) of particles obeys the conservation
law
\beq
\label{numbcons}
{\partial \phi \over \partial t} = - {\partial j \over \partial z},
\eeq
where the shear-induced current \cite{leigh,phillips} can be written
as
\beq
\label{current}
j = -f_c(\phi) a^2 \gdot {\partial \phi \over \partial z}
- f_s(\phi) a^2 \phi {\partial \gdot \over \partial z},
\eeq
In (\ref{current}), $a$ is the particle radius, and  $f_c$ and
$f_s$ are dimensionless functions of the particle volume fraction
$\phi$ \cite{leigh,phillips,nofree}. Note that (\ref{current})
says that particles can move in the absence of concentration
gradients, or even {\em against} concentration gradients, if the
gradient in the deformation-rate is appropriately directed. These
equations are an essential ingredient of our theory of the
shear-induced segregation seen in \cite{acr1,tiru}.

The main result of our analysis is that equations (\ref{numbcons}) and (\ref{current}),
when applied to neutrally buoyant Stokesian suspensions in horizontal rotating
cylinders, predict precisely the instability seen in the experiments of
\cite{acr1,tiru}, if the concentration is large enough. The growth
rate $\Gamma_q$ of
the instability varies as $q^2$ for small wavenumber $q$.
The parameters which govern the instability depend only on the volume
fraction $\phi$ of the suspension and the fill fraction of the cylinder.
At low rotation rates, $\Omega$ scales completely
out of the problem: the  range of unstable
 wavenumbers is therefore independent of $\Omega$.
For $\phi$ just above the instability-onset value $\phi_c$,
$\Gamma_q$ reaches a maximum at $q = q_* \sim \sqrt{(\phi -
\phi_c) \rho g /\sigma}$ where $\sigma$ is the surface tension of
the suspension. For reasonable values of these parameters we find
that the fastest growing mode has wavelength of order centimetres.
We also explain why the instability disappears for large rotation rates. 

\begin{figure}
\begin{center}
\includegraphics{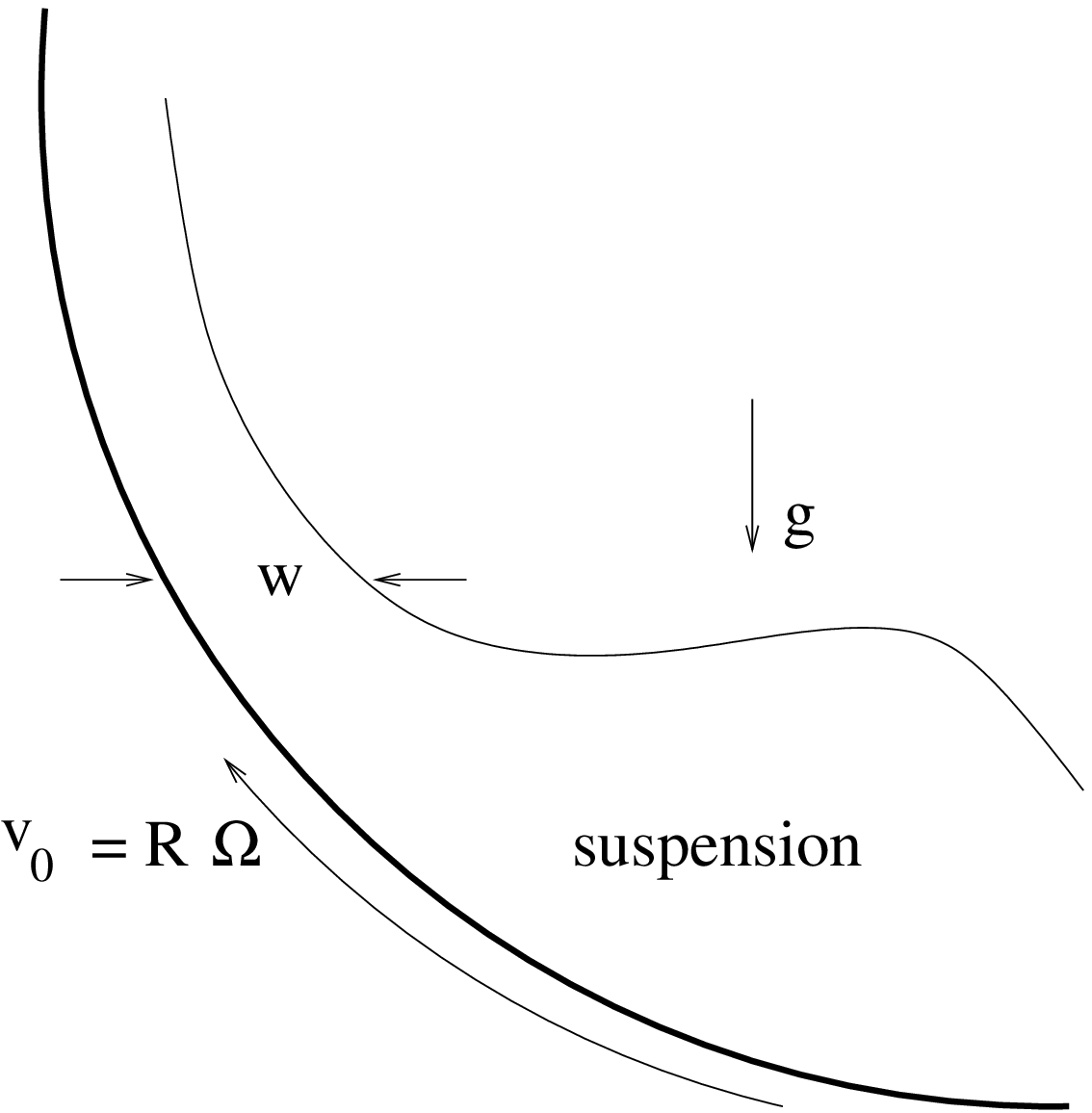}
\end{center}
\caption{\label{fig:onewall} 
}
\end{figure}

We now obtain coupled equations of motion for the particle
concentration and free-surface profile, and show that these lead
naturally to the above results.  Consider a homogeneous suspension with
kinematic viscosity $\nu$, filling a fraction $F$ of the volume of 
a horizontal cylinder
of radius $R$, rotating about its symmetry axis with a tangential
velocity $v_0 = R \Omega$ (See Fig. \ref{fig:onewall}).   
Recall first the results of \cite{tirubump}. 
(a) The dimensionless combination $\beta \equiv F\sqrt{gR^2/\nu v_0}$, 
where $g$ is the acceleration due to gravity, measures the relative importance 
of gravitational and viscous forces;  
(b) As $v_0$ is increased (i.e., as $\beta$ is decreased), a fluid film of 
thickness $\bar{w}$ is dragged up and coats the cylinder wall; (c) Since $\bar{w}$, 
for low speeds, is smaller than the depth of the residual pool of fluid at the bottom 
of the cylinder, the thickness profile has a ``bump'' at the bottom. 
(d) Once $\bar{w}$ reaches a value $\simeq FR$, which occurs for  
$\beta = \beta_c \simeq 1.4$, the growth of $\bar{w}$  
saturates since all the available fluid then coats the cylinder 
more or less uniformly, and the bump disappears.  
(e) For higher speeds, i.e., for $\beta < \beta_c$, 
$\bar{w}$ is effectively independent of the rotation speed, and 
is determined simply by the geometrical statement $\bar{w}/R = F$. 
The mechanism we propose below for the instability applies only when the 
thickness is determined 
by the rotation speed by an explicit balance between viscous and gravitational 
forces which is why, in \cite{tiru}, the instability 
disappears when the bump does. 

Consider a general situation (Fig. \ref{fig:onewall}) where the thickness $w(z,t)$ 
of the fluid film dragged up, 
as well as the volume fraction field $\phi(z,t)$ (and hence the
viscosity), are varying in space and time. The component of the
deformation rate that could vary in the axial direction is given by the
velocity difference across the layer divided by the thickness:
\be
\label{locgdot}
\gdot(z,t) \propto {v_0 - \alpha_2 \dot{w} \over w},
\ee
where $\alpha_2$ is a pure number of order unity,
independent of material parameters.
The experiments of \cite{acr1,tiru} are performed on highly viscous fluids, so that
 the Reynolds number is very small over the entire range of speeds and length scales
studied. We shall therefore work in the limit of zero Reynolds number, where the
inertia of particles and fluid are ignored. Accordingly, the balance of
gravitational, viscous and interfacial forces per unit area of the
layer tells us that
\be
\label{balance1}
\rho g w(z,t)   = \alpha_1 \eta(\phi) {v_0 - \alpha_2 \dot{w} \over w} +
\sigma {\partial^2 w \over \partial z^2}
\ee
and in particular that the layer thickness in the
{\em steady, spatially uniform} state is
\be
\bar w   = \sqrt{\alpha_1 \eta(\phi) v_0 \over \rho g}.
\label{balance2}
\ee
In (\ref{balance1}) and (\ref{balance2}), $\rho$, $\eta(\phi)$, and $\sigma$ are
 respectively the density, effective viscosity (as a function \cite{einstein,krieger} of the local particle volume fraction
$\phi$) and surface tension of the suspension, $g$ is the acceleration due to gravity,
and $\alpha_1$ is another geometrical factor of order unity, with no dependence on any
material parameter. The fill fraction $F$ determines the angle made by the
cylinder with the free surface of the pool of suspension, and hence the 
details of the flow in the fluid layer. This is reflected in the parameters 
$\alpha_1$ and $\alpha_2$ in our model, but our conclusions are qualitatively 
insensitive to their precise numerical values.

Let us now perturb the thickness and concentration fields:
$[w(z,t),\, \phi(z,t)] = [\bar{w} + \delta w (z,t), \, \phi_0 + \psi(z,t)]$. This will
in turn lead to perturbations of the local values of $\gdot$ and $\eta$, yielding
closed equations of motion for the evolution of $\delta w (z,t)$ and
$\psi(z,t)$ via (\ref{numbcons}) and (\ref{balance1}). We work to linear order in
$\delta w$ and $\psi$. Let us work in terms of the nondimensional quantities $H
\equiv \delta w / \bar{w}, \, \tau \equiv (v_0 / \alpha_2 \bar{w})t, \, \zeta
\equiv z / a$. $\psi$ is of course already dimensionless.
Note that to write (\ref{numbcons}) and (\ref{balance1})
in terms of thickness and concentration
fluctuations, we must use (\ref{locgdot}) to express the local deformation-rate in
(\ref{current}) in terms of the thickness perturbation, and replace a local viscosity
perturbation by a local concentration fluctuation via
$\delta \eta / \eta  \simeq {\cal N} \psi$ where
\be
\label{caln}
{\cal N} \equiv {\partial \ln \eta \over \partial \phi}(\phi = \phi_0).
\ee
Although the procedure is straightforward,
some care must be taken in obtaining the perturbation equation from
(\ref{numbcons}): the perturbation of $\gdot$ will involve $\dot{w}$, which
must then once again be eliminated in favour of $\delta w, \, \psi$.
Carrying out these steps, and Fourier-transforming with respect to $\zeta$, i.e.,
considering spatial variation of the form $\exp i q \zeta$,
we find that the Fourier components $H_q$, $\psi_q$ obey
\bea
\label{eom}
 & \pder{}{t}& \left[\matrix{H_q \cr \psi_q}\right] =
\mob \left[\matrix{H_q \cr \psi_q}\right] \nonumber \\
&\equiv&
\left[\matrix{ -(2 + \Sigma q^2)  & {\cal N}  \cr -Sq^2(1 + \Sigma q^2)  & - (C - {\cal
N}S)q^2 }\right] \left[\matrix{H_q \cr \psi_q}\right],
\eea
where
\beq
\label{nondims}
S \equiv \alpha_2 \phi_0 f_s(\phi_0), \, \,  C \equiv \alpha_2 f_c(\phi_0) \, \,
{\rm and}  \, \,  \Sigma \equiv {\sigma \over \rho g a^2}.
\eeq
The stability or otherwise of our sheared suspension is determined by
the characteristic equation
\beq
\label{charac}
\lambda^2 + (2 + Dq^2) \lambda + E q^2 + C \Sigma q^4 = 0
\eeq
for the eigenvalues $\lambda$ of the dynamical matrix $\mob$ in (\ref{eom}),
where
\bea
\label{DE}
D &\equiv& \Sigma + C - {\cal N} S, \nonumber \\
E &\equiv& 2C - {\cal N} S.
\eea
For $q \to 0$, the solutions to (\ref{charac}) are
\begin{mathletters}
\begin{eqnarray}
\label{eig1}
\lambda_1 &\simeq& -{E \over 2} q^2,
\\
\nonumber
\\
\lambda_2 &\simeq& -2 + ({E \over 2} - D) q^2.
\label{eig2}
\end{eqnarray}
\label{eig}
\end{mathletters}
We see from (\ref{eig1}) that the uniform state is linearly unstable to
segregation and thickness modulation if $E < 0$,
with perturbations growing at a rate $\Gamma_q \sim q^2$ at small $q$, and particles
concentrating in the thick regions.
It is straightforward to show
that $\lambda_1$ turns over at larger $q$,
passing through zero at $q = \sqrt{-E/C\Sigma}$, with a peak located,
for $E \to 0^-$, at
\beq
\label{qstar}
q_* \simeq \sqrt{-E/2 C \Sigma}
\eeq
which determines the observed wavelength of the initial instability \cite{foot2}.
For a given fill
fraction, $E$ can vary only with the volume fraction $\phi$. If there is an
instability, it must then be because $E$ turns negative,
in general as $\phi_c - \phi$, as $\phi$ crosses a critical
value $\phi_c$.
This leads to the main result presented at the start of the paper.
In terms of the parameters in (\ref{current}) and (\ref{caln}),
the instability criterion is
\beq
\label{criterion}
\left[{\phi \over \eta} {\partial \eta \over \partial \phi}
\right]_{\phi_0} > {2 f_c(\phi_0) \over f_s(\phi_0)},
\eeq
which should in general happen in real suspensions at large enough $\phi_0$.

In more detail, note that the coefficient $f_c(\phi)$ originates \cite{leigh,phillips}
from a direct shear-induced self-diffusion as well as from a tendency to move down
{\em viscosity} gradients. The latter tendency opposes the instability, as we shall
now show. For a Newtonian suspension, $\eta$ varies only if
$\phi$ does. Thus, by the arguments of Leighton and Acrivos \cite{leigh},
we can write the current in (\ref{numbcons}) as
\beq
\label{currphil}
j = -a^2 \phi [M_s \phi {\partial \gdot \over \partial z} +
\gdot (M_c  + M_{\eta}{\cal N}) {\partial \phi \over \partial z}],
\eeq
where $M_c$, $M_s$ and $M_{\eta}$ are order-unity phenomenological quantities.
Comparing (\ref{currphil}) and (\ref{current}) we see that $f_s = \phi M_s$
and $f_c = \phi(M_c + {\cal N} M_{\eta})$, and the instability criterion
(\ref{criterion}) becomes
$(M_s - 2 M_{\eta}) {\cal N} > 2 M_c$.
 The experiments of \cite{leigh,phillips} have determined these
 coefficients for diffusion in the gradient direction, while diffusion
 in the problem we consider is in the vorticity direction. Further, a
 microscopic theory for determining them is also not available.
 Provided $M_s > 2 M_{\eta}$, the
 growth of the viscosity with $\phi$ \cite{einstein,krieger} means
 that the instability should always arise at large enough
 concentration. An independent measurement of these coefficients is clearly
 called for.

We now assume that the uniform state is unstable and ask whether the typical
wavenumber of the segregation changes from its initial value $q_*$, and
whether nonlinear terms cause the exponential growth to saturate at long times.
It suffices to look at
the dynamics for the slow mode (eigenvalue $\lambda_1$  in the linearised
limit). In that case $H_q$ in (\ref{eom}) is ``slaved'' to $\psi_q$, $H_q \simeq
-{\cal
N}\psi_q/(2 + \Sigma q^2)$. Expanding (\ref{current}) about
$\phi_0$, and retaining terms nonlinear in $\psi_q$, we obtain the effective equation
of motion
\beq
\label{cahnhil}
{\partial \psi \over \partial t} = {\partial^2 \over \partial z^2}
({E \over 2} \psi  + c_2\psi^2 + c_3 \psi^3 .....)
- {{\cal N} S \over 4} {\partial^4 \psi \over \partial z^4}
\eeq
where $c_2$ and $c_3$ are coefficients arising from the $\phi$ dependence of $f_c$,
$f_s$, and $\eta$. Eqn. (\ref{cahnhil}) is well-known in the domain-growth literature
\cite{bray,satya}. In particular, it has been shown \cite{kawa} that the nonlinear terms
in (\ref{cahnhil}) with $E < 0$ cause the characteristic wavelength $L(t)$ of the
pattern of segregation at time $t$ to grow as $\ln t$ at long times. This extremely slow
growth should in principle be testable by a patient experimenter.

    We have shown that two classic properties of non-Brownian
suspensions, {\it viz.}, concentration-dependent viscosity
\cite{einstein,krieger} and shear-induced diffusion \cite{leigh},
lead to a natural explanation of the experiments of
\cite{acr1,tiru} on segregation in suspensions in partially
filled, rotating horizontal cylinders. Our dynamical equations, at
high enough concentration, display an instability towards axial
segregation and a modulation of the free surface, with  particles
accumulating under the crests of the modulation. For parameter
values, say, 10$\,$\% past the instability, taking a plausible
surface tension of $30\,$dyne/cm, (\ref{qstar}) implies a
wavelength of about $3\,$cm for the fastest growing mode at onset,
which is consistent with the experiments of \cite{tiru}.
Independent measurements of the parameters in (\ref{criterion})
and (\ref{qstar}), in transient experiments, should provide a
stringent test of our theory, as should studies of the long-time
behaviour of the wavelength of the segregation pattern.


\newpage
\centerline{\bf FIGURE CAPTIONS}
Figure 1:
Schematic of steady-state surface profile of a
horizontal cylinder of radius $R$, rotating with angular speed $\Omega$.
The thickness $w$ of the suspension layer dragged up, as well the concentration
of solute indicated by shading, are modulated with respect to the axial coordinate $z$
(see ref. \cite{tiru}).

Figure 2:
Cross-sectional schematic of profile of fluid layer
of characteristic thickness $w$ and kinematic viscosity $\nu$, dragged
up against gravity $g$ by the wall of a cylinder, rising at tangential 
speed $v_0$.

\newpage


\newpage

\begin{thebibliography}
\bibitem{foot2}
The second eigenvalue $\lambda_2$ could also conceivably turn positive at
much larger ${\cal N}S$, which would give an instability of a different
character, with onset at a nonzero wavenumber. We shall focus here on the
behaviour of $\lambda_1$.
\end{thebibliography}
\end{document}